\documentclass[a4paper,11pt]{article}
\pdfoutput=1 
\usepackage{jcappub}
\usepackage[T1]{fontenc} 
\makeatletter
\def\@fpheader{\relax}
\makeatother
\usepackage{amsmath}
\usepackage{array}
\usepackage{caption}
\usepackage{geometry}
\usepackage{colortbl} 
\usepackage[dvipsnames]{xcolor}
\usepackage{soul} 
\usepackage{cancel}
\usepackage{hyperref}
\usepackage{url}

\definecolor{Green}{rgb}{0,0.5,0}

\usepackage{subcaption}

\title{Constraining Effective Field Theories for dark matter candidates annihilating into gamma-ray lines with CTAO}

\author[a,b,c]{Lucia Angel}
\author[d]{, Debasish Borah}
\author[a,b,e,f]{, Jacinto P. Neto}
\author[a,b,c,g]{, Farinaldo S. Queiroz}
\author[h]{, Vitor de Souza}

\affiliation[a]{Departamento de F\'isica, Universidade Federal do Rio Grande do Norte, 59078-970, Natal, RN, Brasil}
\affiliation[b]{International Institute of Physics, Federal University of Rio Grande do Norte, Campus Universit\'ario, Lagoa Nova, Natal-RN 59078-970, Brazil} 
\affiliation[c]{Millennium Institute for Subatomic Physics at the High-Energy Frontier (SAPHIR) of ANID, Fern\'andez Concha 700, Santiago, Chile}
\affiliation[d]{Department of Physics, Indian Institute of Technology, Guwahati, Assam 781039, India}
\affiliation[e]{Dipartimento di Scienze Matematiche e Informatiche, Scienze Fisiche e Scienze della Terra, Universita degli Studi di Messina, Viale Ferdinando Stagno d’Alcontres 31, I-98166 Messina, Italy}
\affiliation[f]{Max Planck Institut für Kernphysik, Saupfercheckweg 1, D-69117 Heidelberg, Germany}
\affiliation[g]{Universidad de La Serena, Casilla 554, La Serena, Chile}
\affiliation[h]{Instituto de Física de São Carlos, Universidade de São Paulo, Av. Trabalhador São Carlense 400, São Carlos-SP, 13566-590, Brazil}

\emailAdd{lucia.correa.717@ufrn.edu.br}
\emailAdd{dborah@iitg.ac.in}
\emailAdd{jacinto.neto.100@ufrn.edu.br}
\emailAdd{farinaldo.queiroz@ufrn.br}
\emailAdd{vitor@ifsc.usp.br}

\abstract{Gamma-ray lines constitute a smoking gun signature
for annihilating dark matter particles. Imaging Atmospheric Cherenkov Telescopes and satellites have searched for such signals but null results have been reported thus far. We take advantage of the expected gamma-ray flux sensitivity of the Cherenkov Telescope Array Observatory (CTAO) toward the direction of the Galactic Centre and Dwarf Galaxies and its exquisite energy resolution to derive upper limits on fermionic and scalar dark matter annihilations into gamma-ray lines. We consider the lowest-order effective operators for scalar and fermion dark matter, and derive limits on the energy scale using the recent CTAO projected sensitivity. Putting our findings into perspective with existing limits from direct and indirect detection experiments, we conclude that CTAO will either play a complementary role or be a discovery channel for dark matter signals.}

\begin{document} 
\maketitle
\flushbottom

\section{Introduction}
\label{sec:intro}

The nature of dark matter (DM), despite numerous cosmological and astronomical observations, remains elusive \cite{Planck:2018vyg}. Various experimental efforts aim to detect non-gravitational interactions of dark matter, including direct detection through nuclear scattering \cite{LZCollaboration:2024lux, XENON:2023cxc} and production in particle accelerators \cite{ATLAS:2024kpy, CMS:2024zqs}. An exciting approach involves identifying dark matter annihilation products into Standard Model (SM) particles \cite{Fermi-LAT:2015kyq, HESS:2018cbt, Foster:2022nva, CTAO:2024wvb}, known in the literature as indirect dark matter detection. Among the possible annihilation products, gamma rays and neutrinos are excellent messengers as they are not deflected by magnetic fields, thus retaining directional information. Unlike charged particles, they travel largely unimpeded, allowing them to carry spectral and spatial information across the cosmos.

It is plausible that dark matter particles, which interact with the SM particles in the early Universe, continue to do so today, particularly in regions with high dark matter density such as the Galactic Centre (GC) and Dwarf Spheroidal Galaxies (dSphs). The GC has the largest $J$-factors, making it the brightest dark matter source in the sky and, for this reason, a typical target for dark matter signals. On the other hand, dSphs, which are dominated by dark matter with little interstellar gas, do not have as large a J-factor, but are subject to relatively fewer background sources \cite{Bringmann:2012ez, Gaskins:2016cha, PerezdelosHeros:2020qyt}. In other words, the GC often gives rise to the strongest limits. 

These annihilations can generate a continuous gamma-ray emission with a bump-like feature and are subject to challenging astrophysical background sources \cite{Abazajian:2010zb, Abazajian:2010sq, Bringmann:2011ye, Abazajian:2011tk, Abazajian:2011ak, Fermi-LAT:2015att, Baring:2015sza, Garcia-Cely:2013zga, Queiroz:2016zwd, Profumo:2016idl, HESS:2016mib, Profumo:2017obk, Queiroz:2019acr, Abazajian:2020tww, Siqueira:2021lqj, Bose:2022ghr}. However, the annihilation into photon pairs, resulting in monochromatic photons, provides a distinct signature due to the relative scarcity of similar astrophysical phenomena. Detection of such gamma-ray lines could directly point to the dark matter mass because the spectrum has a Gaussian-like shape centered at the dark matter mass. Gamma-ray observatories such as H.E.S.S and {\it Fermi} LAT have searched for gamma-ray line signatures from the Milky Way's central region but found no excess signal, setting upper limits on the dark matter annihilation cross-section \cite{Fermi-LAT:2015kyq,HESS:2018cbt, Foster:2022nva}. 

In models where the dark matter particle is thermally produced, we expect an annihilation cross-section around $\langle \sigma_{\rm ann} v\rangle \sim 10^{-26}\,{\rm cm^{3}\,s^{-1}}$ \cite{Arcadi:2024ukq}. Current gamma-ray telescopes are now probing this thermal cross-section in the $10$~GeV to $100$~GeV mass range for particular annihilation final states. Dark matter annihilations into fermions and gauge bosons produce generic spectra, but as the annihilation cross-section is typically orders of magnitude larger than that into gamma-ray lines, they offer a better constraining power. Although, gamma-ray lines are loop-suppressed processes, with cross-section ranging from $10^{-28}\,{\rm cm^{3}\,s^{-1}}$ to $10^{-31}\,{\rm cm^{3}\,s^{-1}}$ depending on theoretical modeling, they do offer a more promising discovery channel.

We are commencing an era where the detection of positive gamma-ray line signals from TeV dark matter annihilation is within reach. The Cherenkov Telescope Array Observatory (CTAO) is under construction and will outperform the existing gamma-ray instruments. The energy resolution\footnote{
The energy resolution is quantified through the distribution of the relative energy error, $(E_R - E_T/E_T)$, commonly expressed as $\Delta E/E$, where $E_R$ and $E_T$ denote the reconstructed and true energies, respectively, of gamma-ray events observed with CTAO~\cite{CTAOPerformance}.} is smaller than $\Delta E/E <  0.1$, and an effective area~\cite{CTAOPerformance} nearly an order of magnitude larger than that of {\it Fermi} LAT for gamma-ray energies above $1$~TeV make CTAO an excellent laboratory for probing dark matter annihilations and decays. When operating, CTAO shall be able to detect gamma rays from below $20$~GeV to $300$~TeV~\cite{CTAOPerformance}. Therefore, CTAO will effectively probe Weakly Interacting Massive Particles (WIMPs) across three orders of magnitude in energy. 

In this work, we are focused on gamma-ray lines, where energy resolution is the key. Box-shaped signals \cite{Ibarra:2013eda, Ibarra:2015tya} and other enhanced gamma-ray emission near the dark matter mass may mimic the effect of a gamma-ray line \cite{Bringmann:2007nk}. Hence, having a good energy resolution will allow us to discriminate dark matter signals from background sources and give insights into the underlying dark matter model behind the detected signal. The spatial morphology is key to separating a dark matter signal from a background source \cite{CTA:2020qlo,Bringmann:2012ez}. Instead, gamma-ray lines offer more distinct signals hardly mimicked by astrophysical sources, see Fig.~\ref{fig:example} below.

\begin{figure}[h!]
    \centering
    \includegraphics[width=0.9\linewidth]{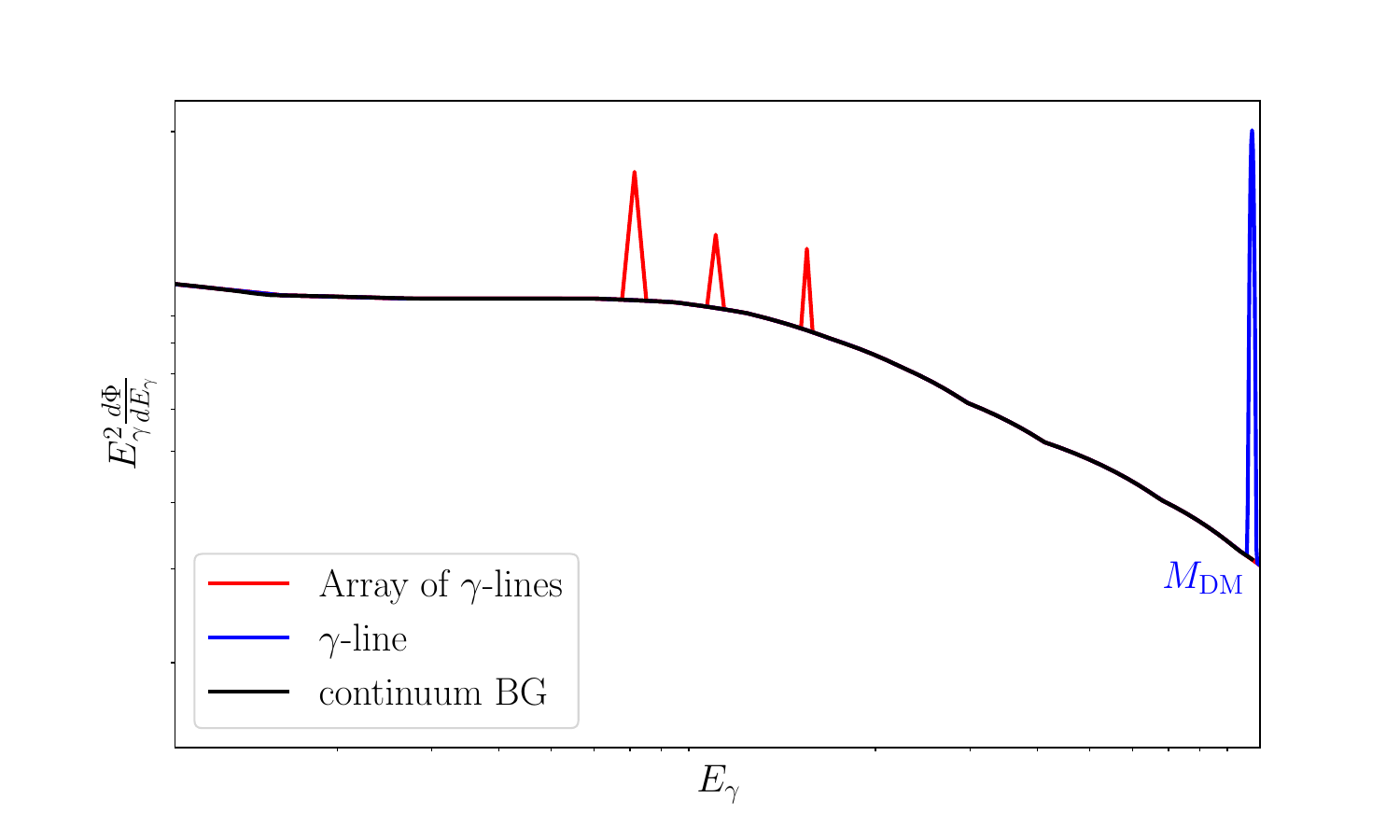}
    \caption{An illustration of how we expect to see gamma-ray lines from the differential flux. Either in a series of lines from a dark matter bound-state model (\textcolor{red}{red}) or a characteristic signal at the end of the spectrum from a model with a specific dark matter mass (\textcolor{blue}{blue}) continuum emissions (black). Figure inspired by \cite{Panci:2024oqc}.}
    \label{fig:example}
\end{figure}

It is worth highlighting that the expected sensitivity we use to constrain the dark matter effective field theory (EFT) considered here is based on the so-called `Alpha' configuration, which is the first-stage of CTAO construction \cite{CTAO:2024wvb, Dubos:2024idb}. Furthermore, the CTAO will achieve better sensitivities than current Imaging Atmospheric Cherenkov Telescopes (IACT) generation instruments by a factor of $5$ to $10$ \cite{CTA2017APh.93.76H}, resulting in an energy resolution of approximately $\Delta E/E \sim \mathcal{O}(0.1)$~TeV. For more details, the interested reader can look at Fig. 1 in \cite{CTAO:2024wvb}. Once again, it shows how powerful and excellent such an instrument is for searching for exotic localized spectral features over many orders of magnitude in gamma-ray energies.

Our goal is to assess the CTAO sensitivity to gamma-ray lines arising from fermion and scalar dark matter stemming from the GC and dSphs. We describe them in terms of effective operators and determine the dark matter signal as a function of the effective energy scale. More importantly than simply showing the CTAO bounds, we put our findings into perspective by comparing them with direct detection experiments and continuum gamma-ray emission stemming from these operators. Therefore, our work advances in comparison to previous assessments of this nature \cite{Cerdeno:2015jca,Duerr:2015wfa,Boddy:2016fds,Profumo:2016idl,Queiroz:2018utk,Yang:2020vxl,Angel:2023rdd,DeLaTorreLuque:2023fyg,Cheng:2023chi}. 

This paper is organized as follows: In Section~\ref{sec1}, we briefly discuss the gamma-ray line from DM annihilation followed by listing the relevant EFT operators for DM annihilation into gamma-rays in Section~\ref{sec:ops}; In Section~\ref{sec:sigma}, we calculate DM annihilation into two photons for both scalar and fermion DM considering dimension six and dimension five operators respectively; In Section~\ref{sec:limits} we discuss the existing bounds on such DM EFT operators from continuum gamma-ray emission and direct detection; Lastly, in Section~\ref{sec:results} we present our results before drawing our conclusions in Section~\ref{sec:conclude}.

\section{Signal from the Dark Side}
\label{sec1}
We devote this section to introducing essential definitions related to the gamma-ray flux. The differential gamma-ray flux observed in a given solid angle element $d\Omega$ on the sky for the $X \overline{X} \to \gamma\gamma$ annihilation process, with photon energy $E = m_X$, is given by~\cite{Gaskins:2016cha,Cirelli:2024ssz}
\begin{equation}
    \frac{d\Phi}{d\Omega \,dE} = \frac{r_\odot}{4\pi} \left(\frac{\rho_\odot}{2 \, m_{X}}\right)^2  J_{\rm ann}(\theta)\,\langle \sigma_{\rm ann} v \rangle \,\frac{d N_{\gamma}}{dE} 
\end{equation} where $m_X$ is the mass of a generic dark matter particle $X$, which we consider not to be its own antiparticle, $\rho_\odot$ is the dark matter energy density at the position, $r_\odot$, of the Solar System in the Milky Way, $\langle \sigma_{\rm ann} v \rangle$ is the velocity averaged annihilation cross-section, with $v$ standing for the dark matter relative velocity, and the annihilation $J_{\rm ann}$-factor, integrated over the line-of-sight ({\rm l.o.s.}),  is defined as,
\begin{equation}
    J_{\rm ann} (\theta) = \int_{\rm l.o.s.} \left(\frac{\rho(r)}{\rho_\odot}\right)^2\, \frac{dl}{r_\odot},
\end{equation}with $\rho(r)$ being the dark matter density profile, located at a distance $l$ to the Earth and angle $\theta$ concerning the center of the dark matter halo, so that $r(l,\theta)$. Finally, the total gamma-ray energy spectrum per annihilation is,
\begin{equation}
    \frac{d N_{\gamma}}{dE} = 2 \delta(E - m_X),
\end{equation} where we get a factor of 2 since each annihilation produces two photons with energy equal to the dark matter particle mass.

\begin{figure}[t]
  \centering
  \includegraphics[width=0.7\linewidth]{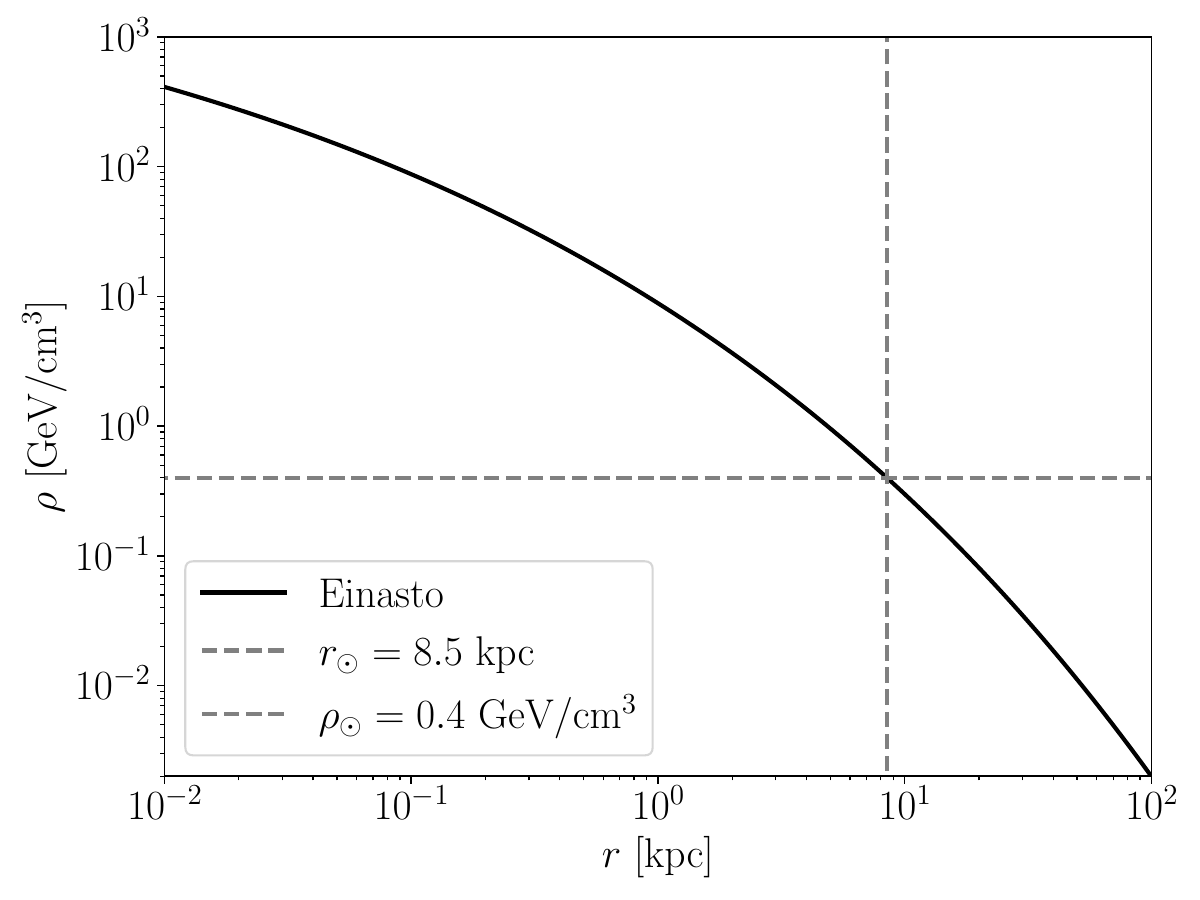}%scale=0.5
  \caption{Einasto dark matter density profiles for the Galactic halo as a function of the distance to the Galactic Centre.}
  \label{fig:prof}
\end{figure}

It is important to note that $\Omega$ represents the solid angle in the sky and is used in the differential flux, while $\theta$ describes the observation direction relative to a point of interest (such as the center of a dark matter halo) and is used in calculating the $J$-factor. By considering both $\theta$ and $\Omega$, we can accurately describe the spatial distribution of the gamma-ray flux from dark matter annihilation across the sky and understand how it varies based on the observation direction.

The predicted dark matter signal from the Galactic Centre is subject to large uncertainties from the astrophysical J-factor, as the dark matter density profile in the inner Galaxy is poorly constrained by observations. Guided by N-body simulations, the CTAO collaboration has adopted the Einasto profile as its benchmark (see Fig.~\ref{fig:prof})~\cite{CTAO:2024wvb, CTA:2020qlo}, justifying our choice of adopting the same profile.

\section{Dark Matter Effective Field Theory}
\label{sec:ops}

We explore an EFT description of the dark matter annihilation into gamma-ray lines. This is a minimalistic approach as the DM phenomenology can be described with very few free parameters, namely, DM mass, cutoff scale, and the corresponding Wilson coefficients. Such EFT descriptions of DM have been studied extensively in the context of direct detection, indirect detection, as well as collider searches in several works \cite{Beltran:2008xg, Fan:2010gt, Goodman:2010ku, Beltran:2010ww, Fitzpatrick:2012ix, GAMBIT:2021rlp}, also summarised in a recent review \cite{Bhattacharya:2021edh}. In this present framework, we consider two possible DM candidates namely, a complex scalar $S$ and a Dirac fermion $\chi$ which are singlets under the $SU(3)_c \otimes SU(2)_{L} \otimes U(1)_{Y}$ gauge symmetry of the SM. A separate symmetry ensures the stability of DM such that all effective operators involve DM bilinears only \footnote{We remain agnostic about the possible UV completions of such stabilising symmetry.}. We focus on effective operators of the lowest order because higher-order operators are relatively suppressed by higher powers of the effective energy scale $\Lambda$. 

There are inherent advantages and limitations to the EFT approach. It is particularly applicable when the momentum transfer in a process is significantly lower than the mass of the mediator linking the dark and visible sectors. For dark matter annihilations, which are typically non-relativistic, the momentum transfer is of the order of the dark matter mass, making EFT a convenient method to evaluate such models. In Tables~\ref{tab:Operatorlist-scalar} and \ref{tab:Operatorlist-fermion}, we list the lowest-order effective operators that produce gamma-ray lines for both scalar and fermionic dark matter types, consistent with the chosen symmetries of our framework.

Our EFT framework includes the SM particles plus a dark matter component, $X$, which can be either a complex scalar or a Dirac fermion. If $X$ is an antiparticle of itself, the results differ by a factor of two. As our goal is to study DM annihilation into photons, the EFT is formulated in terms of DM bilinears and $W^{a}_{\mu\nu}$, $B_{\mu\nu}$ representing the field strength tensors for the $SU(2)_L$ and $U(1)_Y$ groups, respectively\footnote{A general field strength tensor is defined as $F^{a}_{\mu\nu} \equiv \partial_\mu F^{a}_\nu - \partial_\nu F^{a}_\mu + g f^{abc}F^{b}_\mu F^{c}_\nu$, where $g$ is the coupling constant and $f^{abc}$ are the structure constants for the symmetry group, with $f^{abc} = 0$ for Abelian theories.}. The operators listed in Tables~\ref{tab:Operatorlist-scalar}-\ref{tab:Operatorlist-fermion} are expressed in terms of these field strength tensors of gauge fields, which, after spontaneous electroweak symmetry breaking, relate to the mass eigenstates,
\begin{eqnarray}
    B_\mu &= \cos\theta_W A_\mu  - \sin\theta_W Z_\mu, \\ 
    W^{3}_\mu &= \sin\theta_W A_\mu + \cos\theta_W Z_\mu, 
\end{eqnarray}
where $A_{\mu}$ and $Z_\mu,$ are the neutral electroweak gauge bosons mass eigenstates, and $\theta_W$ is the Weinberg angle.  Our analysis will focus exclusively on effective operators up to dimension six, as these encompass all potential leading-order annihilation channels into gamma-ray lines for fermion and scalar DM. This comprehensive list forms the basis of our study, delimiting the scope of these specific interactions.

%%%

\begin{table}[h]
\centering
\begin{tabular}{|>{\centering\arraybackslash}m{1.5cm}|>{\centering\arraybackslash}m{8cm}|>{\centering\arraybackslash}m{3cm}|}
\hline
\multicolumn{3}{|c|}{\cellcolor{blue!20} Complex Scalar Dark Matter -- Dimension 6 Operators} \\
\hline
S1 & $\frac{1}{\Lambda^2_{\rm S1}}\,SS^* B_{\mu\nu}B^{\mu\nu}$ & $\gamma \gamma$ \\ 
\hline
S2 & $\frac{1}{\Lambda^2_{\rm S2}}\,SS^* W^a_{\mu\nu}W^{a\mu\nu}$ & $\gamma \gamma$, $\gamma Z$ \\
\hline
S3 & $\frac{1}{\Lambda^2_{\rm S3}}\,SS^* B_{\mu\nu}\tilde{B}^{\mu\nu}$ & $\gamma \gamma$\\
\hline
S4 & $\frac{1}{\Lambda^2_{\rm S4}}\,SS^* W^a_{\mu\nu}\tilde{W}^{a\mu\nu}$ & $\gamma \gamma$, $\gamma Z$ \\
\hline
\end{tabular}
\caption{List of effective interactions for complex scalar dark matter and the type of line signals ($\gamma \gamma$, $\gamma Z$).}
\label{tab:Operatorlist-scalar}
\end{table}

\begin{table}[h]
\centering
\begin{tabular}{|>{\centering\arraybackslash}m{1.5cm}|>{\centering\arraybackslash}m{6.5cm}|>{\centering\arraybackslash}m{3cm}|}
\hline
\multicolumn{3}{|c|}{\cellcolor{blue!20} Dirac Fermion Dark Matter -- Dimension 5 Operators} \\
\hline
F1 & $\frac{1}{\Lambda_{\rm F1}}\,\bar{\chi}\gamma^{\mu\nu}\chi B_{\mu\nu}$ &  $\gamma \gamma$ \\
\hline
F2 & $\frac{1}{\Lambda_{\rm F2}}\,\bar{\chi}\gamma^{\mu\nu}\chi \tilde{B}_{\mu\nu}$ &  $\gamma \gamma$, $\gamma Z$ \\
\hline
\end{tabular}
\caption{List of effective interactions for Dirac fermion dark matter and the type of line signals ($\gamma \gamma$, $\gamma Z$). Note that $\gamma^{\mu\nu} = \frac{i}{2}[\gamma^\mu, \gamma^\nu]$.}
\label{tab:Operatorlist-fermion}
\end{table}

With these operators at hand, our next step is to compute their averaged annihilation cross-section.

\section{Averaged Annihilation Cross-Sections}
\label{sec:sigma}

The thermally averaged annihilation cross-section normalizes the gamma-ray flux resulting from dark matter annihilation, as can be seen in Eq.\eqref{eqsigmaann}. In the non-relativistic limit, it can be approximated to be  \cite{Gondolo:1990dk}
\begin{equation}
    \langle \sigma_{\rm ann} v \rangle \simeq \sigma_0 + \frac{1}{2}b \,v^2,
    \label{eqsigmaann}
\end{equation}
where $\sigma_{\rm s} \equiv \sigma_0$ is the s-wave contribution that does not depend on the velocity and $\sigma_{\rm p} \equiv b \,v^2/2$ is the velocity-dependent p-wave contribution. Such an expression translates the impact of the velocity dependence of the cross-section, as the dark matter velocity changes from freeze-out to the late-time Universe. We expect dark matter particles to be cold today, with radial velocity dispersion $v \sim \mathcal{O}(10^{-3})$ in the Galactic halo \cite{Slatyer:2021qgc}. Thus, we can safely neglect the p-wave component and focus our reasoning on the s-wave term. 

We assume $p_1$ and $p_2$ as the momenta of the incoming dark matter particles, $p_3$ and $p_4$ as the momenta of the outgoing particles. For concreteness, we take $p_4$ as the momentum for the Z boson for $\gamma Z$ final state.  Hence, the differential cross-section is written,
\begin{equation}
\frac{d\sigma_0}{d\Omega} = \frac{E_3}{256\pi^2~E^3~v} \overline{|{\cal M} |^2},
\label{eqCS}
\end{equation}
where $E = m_X + {\cal O}(v^2)$ is the energy of each dark matter particle, $v$ is the dark matter velocity and $E_3 = |\vec{p}_3|$ is the energy of the outgoing photon. Moreover, $\overline{|{\cal M} |^2}$ is the amplitude ${\cal M}$ squared averaged over initial dark matter spins %(if any) 
and summed over final state particle spins, which will be obtained below. Therefore, once we find $\overline{|{\cal M} |^2}$, we can immediately find the thermally averaged annihilation cross section, $\sigma_0$, using Eq.\eqref{eqsigmaann} and Eq.\eqref{eqCS}.

The effective interactions between the scalar dark matter $S$ and the neutral electroweak gauge bosons are shown in Tab.~\ref{tab:Operatorlist-scalar} via the S1$-$S4 operators. For S1 and S2 operators, involving the contraction of the strength tensors $B_{\mu\nu}B^{\mu\nu}$ and $W^{a}_{\mu\nu}W^{a\mu\nu}$, respectively, the amplitude for two final state photons is,
\begin{equation}
\mathcal{M}({SS^* \rightarrow \gamma \gamma}) = 2 Y \left[ (p_3\cdot p_4)(\varepsilon_3 \cdot \varepsilon_4)- (p_3\cdot \varepsilon_4)(p_4 \cdot \varepsilon_3) \right], \quad {\rm with} \quad Y \equiv \frac{\cos^2 \theta_W}{\Lambda},
\end{equation}
where $p_{3, 4}$ are the outgoing momenta of the photons with respective polarization vectors $\varepsilon_{3,4}$. Summing over the spins, the amplitude squared reads,
\begin{equation}\label{eq:Scalar2Gamma}
\sum_{\varepsilon_e, \varepsilon_4} |\mathcal{M}({SS^* \rightarrow \gamma \gamma })|^2 = 4YY^\dagger \frac{s^2}{2} = \frac{2 s^2 \cos^4\theta_W}{\Lambda^4},
\end{equation}
where $s^2=(p_3+p_4)^2$. Considering the same operators but for a photon and a $Z$ final state, the spin-averaged amplitude squared is,
\begin{equation}\label{eq:ScalarGammaZ}
\sum_{\varepsilon_e, \varepsilon_4} |\mathcal{M}({SS^* \rightarrow \gamma Z})|^2 = 4Y_1Y_1^\dagger \frac{1}{2} \left( s-m_Z^2 \right)^2 = \frac{2  \sin^2 (2 \theta_W)}{\Lambda^4} \left( s- m_Z^2 \right)^2,
\end{equation}where $Y_1=\sin(2\theta_w)/\Lambda$ and $p_4$ the $Z$ boson momentum. Finally, we plug Eqs.~\eqref{eq:Scalar2Gamma} and \eqref{eq:ScalarGammaZ} into Eq.~\eqref{eqCS}, to obtain the average annihilation cross-sections for the S1 (S3) operator into $\gamma\gamma$ and $\gamma Z$ lines,
\begin{eqnarray}
\left< \sigma({S S^* \rightarrow \gamma \gamma}) \, v\right> &=& \frac{\cos^4(\theta_W)}{\pi}  \frac{m_S^2}{\Lambda_{\rm S1}^4}, \label{eq:sv1gg} \\
\left< \sigma({S S^* \rightarrow \gamma Z}) \, v\right> &=& \frac{\sin^2 (2\theta_W)}{\pi}  \frac{m_S^2}{\Lambda_{\rm S1}^4} \left( 1 - \frac{m_Z^2}{4m_S^2}  \right)^3,
\end{eqnarray}
and for the S2 (S4) operator,
\begin{eqnarray}
\left< \sigma({S S^* \rightarrow \gamma \gamma}) \, v\right> &=& \frac{ \sin^4(\theta_W)}{\pi} \frac{m_S^2}{\Lambda_{\rm S2}^4}, \\
\left< \sigma({S S^* \rightarrow \gamma Z}) \, v\right> &=& \frac{\sin^2 (2\theta_W)}{\pi}  \frac{m_S^2}{\Lambda_{\rm S2}^4} \left( 1 - \frac{m_Z^2}{4m_S^2}  \right)^3.
\label{eq:Annscalar}
\end{eqnarray}
The same formalism is applied to the fermionic dark matter $\chi$ for F1 and F2 operators. However, we have to go to second-order perturbation theory to get the relevant contribution for $\gamma\gamma$ and $\gamma Z$ final states. In this case, the averaged annihilation cross-sections into both $\gamma\gamma$ and $\gamma Z$ final states are given by,
\begin{eqnarray}
\left< \sigma ({\chi\bar{\chi} \rightarrow \gamma \gamma}) \,v \right> &=& \frac{8}{\pi} \cos^4(\theta_W) \frac{m_\chi^2}{\Lambda_{\rm F}^4},\\
\left< \sigma ({\chi\bar{\chi} \rightarrow \gamma Z}) \,v \right> &=& \frac{2}{\pi} \sin^2(2\theta_W) \frac{m_\chi^2}{\Lambda_{\rm F}^4} \left(1+\frac{m_Z^2}{4m_\chi^2} \right)^2  \left( 1-\frac{m_Z^2}{4m_\chi^2} \right) . \label{eq:Annfermion}
\end{eqnarray}
Regardless of the astrophysical target, the scalar dark matter annihilation into $\gamma\gamma$ has a larger cross-section for the S1 (associated with $B_{\mu\nu}B^{\mu\nu}$) operator than the S2 (involving $W^{a}_{\mu\nu}W^{a\mu\nu}$) one due to the symmetry factor from the presence of identical particles in the final state. The $\gamma Z$ final state curves slope down as the dark matter mass decreases because the averaged annihilation cross-section is proportional to $1 - m_Z^2/(4 m_S^2)$, which is related to the energy resolution of the experiment as we will discuss below. Comparing the results from the $\gamma \gamma$ and $\gamma Z$ lines for S1 and S2 operators considering $m_S \gtrsim 100$~GeV, we note that for S1 they are very close, but for S2 operator the $\gamma Z$ lines correspond to larger cross-sections. Moreover, comparing Eqs.~\eqref{eq:Annscalar} and \eqref{eq:Annfermion}, we note that the fermion averaged annihilation cross-section is larger than the scalar one for the same DM mass. We assume $\Lambda_{F1}=\Lambda_{F2}=\Lambda_F$ throughout.

One should keep in mind that the $\gamma Z$ lines happen as long as $m_X > m_Z/2$, and the total thermally averaged annihilation cross-section for a dark matter particle $X = S, \chi$ must consider both $\gamma \gamma$ and $\gamma Z$ lines,
\begin{equation}
\left< \sigma_{\rm ann} \, v \right> = \left< \sigma ({XX \rightarrow \gamma \gamma}) \,v \right> + \left< \sigma ({XX \rightarrow \gamma Z}) \,v \right>   \Theta \left[ \frac{\Delta E}{E} - \frac{m_Z^2}{4m_X^2} \right],
\label{eq:Eqsv}
\end{equation} 
where $\Theta$ is the Heaviside step function and $\Delta E/E$ is the energy resolution of the instrument, which is energy-dependent. We highlight there is no infinite-energy-resolution telescope. The telescope's energy dispersion, or finite energy resolution, is related to its response function. The expected energy resolution for CTAO is $\Delta E/E \sim 10\%$ between 1 TeV and 100 TeV \cite{CTAO:2024wvb}.
The final state photons have different energies depending on the number of lines, say,
\begin{align}
    E_{\gamma\gamma} & = m_X, \\
    E_{\gamma Z}  &= m_X - \frac{m_Z^2}{4 m_X}. 
\end{align}
We can check in Eq.~\eqref{eq:Eqsv} via the Heaviside function, that the detector will not be able to distinguish between the two possible final states if the relative difference between their energies is up to the energy resolution of the instrument, i.e. $(E_{\gamma\gamma} - E_{\gamma Z})/E_{\gamma\gamma} \leq \Delta E/E$. Therefore, we must insert the CTAO energy resolution into Eq.~\eqref{eq:Eqsv} to derive the correct constraints.

That said, in the next section, we present the existing bounds on the effective operators based on continuous gamma-ray emission and direct detection.

\section{Existing Limits} \label{sec:limits}
Although we focus on placing bounds from gamma-ray lines, the effective operators in this work also yield a continuum gamma-ray spectrum and DM-nucleon recoil at direct detection experiments. Hence, it is important to put our findings into perspective with direct detection, and the natural continuum gamma-ray spectrum whenever present to assess the role of the CTAO regarding probing gamma-ray lines signals from dark matter annihilation.

\subsection{Continuum gamma-ray bounds}

We present the bounds from H.E.S.S. and expected limits from CTAO for dark matter annihilation into charged final states, such as $W^+W^-$ and $\tau^+\tau^-$, in the GC, considering the Einasto density profile \cite{CTA:2020qlo, HESS:2022ygk}. For the continuum spectrum, the H.E.S.S. dataset was collected over 6 years, and after data quality selection, the analysis used a total of 546 hours of live observation time. The CTAO expected limits assume a total observation time of 525 hours. The Fermi-LAT collaboration has adopted a different density profile, namely, Navarro-Frenk-White (NFW) and generalized NFW (gNFW) \cite{Fermi-LAT:2017opo}. Thus, we decided to keep only the H.E.S.S. bound to make a fair comparison. 

\begin{table}[h!]
\centering
\begin{tabular}{|>{\centering\arraybackslash}m{1.5cm}|>{\centering\arraybackslash}m{8cm}|>{\centering\arraybackslash}m{3cm}|}
\hline
\multicolumn{3}{|c|}{\cellcolor{blue!20}Dark Matter Effective Operators vs Continuum Spectrum Processes} \\
\hline
S1 & $\frac{1}{\Lambda^2_{\rm S1}}\,SS^* B_{\mu\nu}B^{\mu\nu}$ & None \\ 
\hline
S2 & $\frac{1}{\Lambda^2_{\rm S2}}\,SS^* W^a_{\mu\nu}W^{a\mu\nu}$ & $W^+ W^-$ \\
\hline
F & $\frac{1}{\Lambda_{\rm F}}\,\bar\chi \gamma^{\mu\nu}\chi B_{\mu\nu}$ & $W^+ W^-$, $f^+ f^-$ \\
\hline
\end{tabular}
\caption{Dark matter effective operators for continuum spectrum signals. The S1 operator does not generate any {\it continuum spectrum final states} we have considered here. The S2 operator generates only $W^+ W^-$ final states, and the fermionic dark matter, F operator, annihilates to both $W^+ W^-$ and $f^+ f^-$, with $f$ standing for any electrically charged SM fermion \cite{Chen:2013gya}.}
\label{tab:Operatorlist-diffuse}
\end{table}

\begin{figure}[t!]
    \centering
    \begin{subfigure}[b]{0.48\linewidth}
        \includegraphics[width=\linewidth]{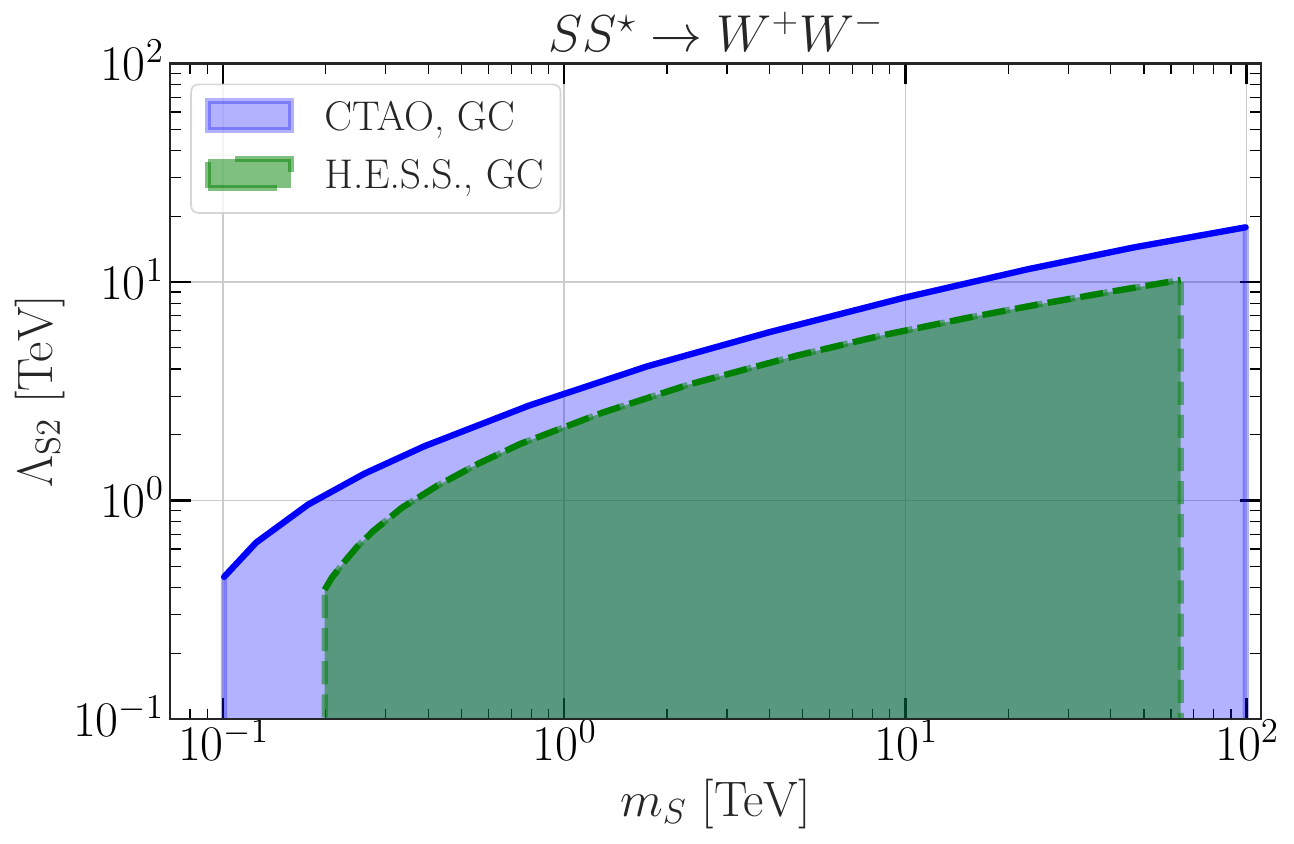}
        \caption{Complex scalar DM}
        \label{fig:lambs2-diffuse}
    \end{subfigure}
    \hfill
    \begin{subfigure}[b]{0.48\linewidth}
        \includegraphics[width=\linewidth]{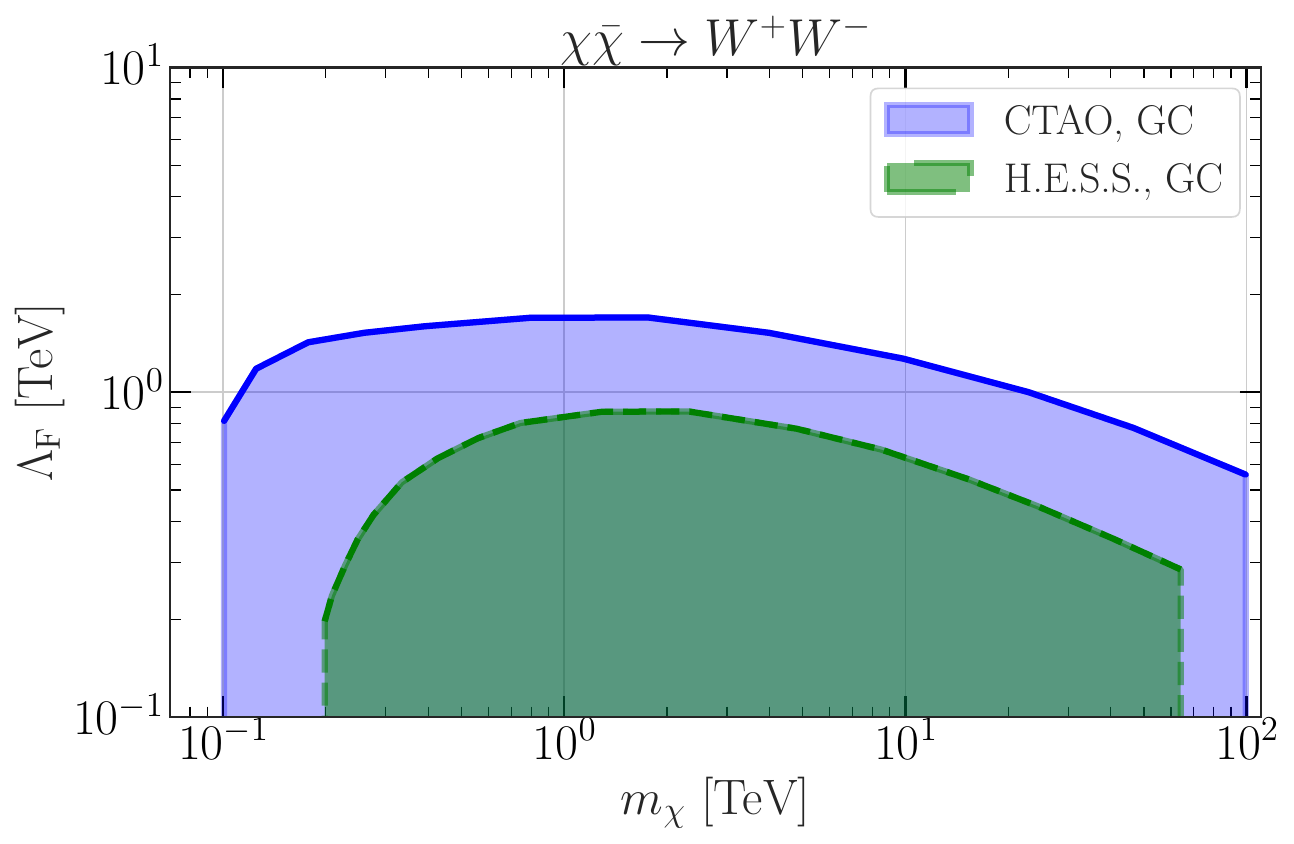}
        \caption{Dirac fermion DM}
        \label{fig:lambf-diffuse}
    \end{subfigure}
    \caption{CTAO (solid blue region) and H.E.S.S. (dashed green region) upper limits on the effective energy scales for the diffuse gamma-rays from  $ X \bar X \to W^{+}W^{-}$ annihilation process. We highlight that the CTAO and H.E.S.S. limits come from GC observations using an Einasto density profile \cite{CTA:2020qlo, HESS:2022ygk}.}
\label{fig:results-diffuse-WW}
\end{figure}

\begin{figure}[t!]
    \centering
    \includegraphics[width=0.5\linewidth]{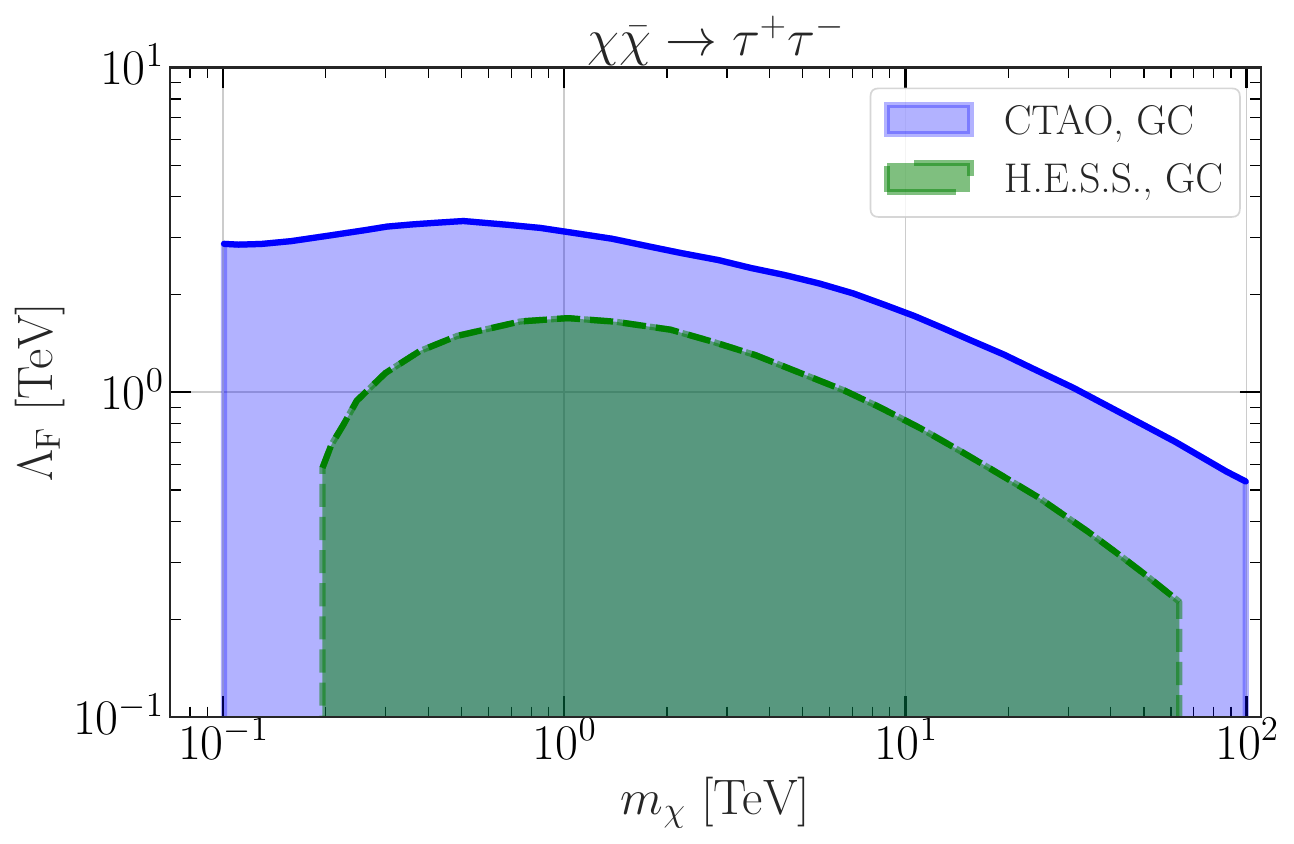}
    \caption{CTAO (solid blue) and H.E.S.S. (dashed green) upper limits on the effective energy scales for the diffuse gamma-rays from  $ X \bar X \to \tau^{+}\tau^{-}$ annihilation process at GC assuming an Einasto density profile.}
\label{fig:results-diffuse-tautau}
\end{figure}

From operator S2, we can have the process $S S^* \rightarrow W^{+} W^{-}$. The thermal cross-section is given by \cite{Chen:2013gya},
\begin{align}
    \langle \sigma (S S^\star \rightarrow W^{+} W^{-}) \,v\rangle = \frac{4}{\pi}\frac{m_S^2}{\Lambda_{\rm S2}^4} \sqrt{1 - \frac{m_W^2}{m_S^2}} \left(1 - \frac{m_W^2}{m_S^2} + \frac{3}{8}\frac{m_W^4}{m_S^4}  \right) 
\end{align}
where $m_W = 80.377$~GeV is $W$ boson mass, and $v_{\rm SM} = 246.22$~GeV is the vacuum expectation value of the SM Higgs field. From operator F directly induces annihilation into all electrically charged pairs of SM particles, but we choose the $W^{+} W^{-}$ and $\tau^{+} \tau^{-}$ channels for exemplification. The corresponding annihilation cross-sections are given by \cite{Chen:2013gya},
\begin{align}
    \langle \sigma (\chi \bar\chi \rightarrow W^{+} W^{-}) \,v\rangle & = \frac{1}{4 \pi} \frac{m_Z^2 \sin^2\theta_W}{v_{\rm SM}^2 \Lambda_{\rm F}^2} \left( 1- \frac{m_Z^2}{4 m_\chi^2} \right)^{-2} \left(1 - \frac{m_W^2}{m_\chi^2} \right)^{3/2} \nonumber \\
    &\times \left( 1 + \frac{23}{4} \frac{m_W^2}{m_\chi^2} + \frac{3}{4}\frac{m_W^4}{m_\chi^4} \right), \\
    \langle \sigma \left(\chi \bar\chi \rightarrow \tau^+ \tau^- \right) v \rangle &= \frac{1}{4 \pi} \sqrt{ 1 - \frac{m_\tau^2}{m_\chi^2}} \frac{m_Z^2 \sin^2\theta_W}{v_{\rm SM}^2 \Lambda^2_{\rm F}} \left( 1 - \frac{m_Z^2}{4 m_\chi^2} \right)^{-2}  \\
    & \times \left[ 4 \mathcal{A}_{\tau}^2 + \left( \frac{m_\tau}{m_\chi}\right)^2 \left( 2 \mathcal{A}_{\tau}^2 - 1 \right) + 1 \right],
\end{align}
where $\mathcal{A}_{\tau} = 2 Q_\tau \left(1 - m_Z^2 / m_\chi^2\right) + 1/2$, and $m_\tau$ and $Q_\tau$ are the $\tau$ mass and electric charge (in natural units). Tab. \ref{tab:Operatorlist-diffuse} summarizes the {\it continuum spectrum charged final states} for each effective operator.

We show the CTAO and H.E.S.S. limits for each final state in Figures \ref{fig:results-diffuse-WW} and \ref{fig:results-diffuse-tautau}. The blue region represents the expected CTAO upper limit for GC. Thus, for both scalar and fermion dark matter CTAO will greatly improve existing limits, reaching effective energy scales above $1$~TeV depending on the dark matter mass and final state. 

It was important to do this exercise because our effective operators also induce a continuous gamma-ray spectrum. Hence, it is clear that CTAO will surpass its predecessors in the detection of a diffuse component gamma-ray component and line emission, as we will address later.

\subsection{Direct detection}

In addition to the existing bounds from the continuum spectrum, we extend the analysis to discuss the approximate direct detection limits on the effective energy scales for the Dirac fermion DM. The operator F can describe the interaction of the Dirac dark matter with photons via an anomalous magnetic dipole moment $\mu_\chi$ \cite{Banks:2010eh, DelNobile:2021wmp},
\begin{eqnarray}
    \mathcal{L}_\chi = - \frac{\mu_\chi}{2}\bar\chi \gamma^{\mu\nu} \chi F_{\mu\nu},
\end{eqnarray}
where $F_{\mu\nu}$ is the electromagnetic field-strength tensor. The interaction between nucleons and photons comes from the SM neutral current Lagrangian,
\begin{eqnarray}
    \mathcal{L}_{\rm NC} \supset - e J^{\mu}_{\rm EM}A_\mu,
\end{eqnarray}
where $e$ is the electric charge and $J^\mu_{\rm EM} = \bar \psi_q \gamma^\mu \psi_q$, for quarks $\psi_q$. In the non-relativistic limit, the DM-nucleus differential scattering cross-section is given by, \cite{Kavanagh:2018xeh,DelNobile:2021wmp} 
\begin{eqnarray}\label{eq:DM-nucleus-diff-cs}
\frac{d\sigma_{\rm T}}{d E_{\rm R}} &=& \frac{e^2 \mu_\chi^2 }{8 \pi} \frac{m_{\rm T}}{m_{\rm N}^2} \frac{1}{v^2} \bigg[ 2 \frac{m_{\rm N}^2}{m_{\rm T}} \bigg( \frac{v^2}{E_{\rm R}} - \frac{m_\chi + 2 m_{\rm T}}{2 m_\chi m_{\rm T}} \bigg)  F_{\rm M}^{p,p}(q^2) + 4 F_\Delta^{p,p}(q^2) \nonumber \\
&-& 2 \sum_{\rm N} g_{\rm N} F_{\Sigma^{\prime}\Delta}^{{\rm N}, N^{\prime}} (q^2) + \frac{1}{4} \sum_{{\rm N, N^\prime}} g_{\rm N} g_{\rm N^\prime} F_{\Sigma^\prime}^{{\rm N, N^\prime}}(q^2) \bigg],
\end{eqnarray}
where $E_{\rm R}$ (expressed in keV) and $q^2 = 2 m_{\rm T} E_{\rm R}$ are the nuclear recoil energy and the recoil momentum, respectively. The relative velocity is $v$, $m_{\rm T}$ denotes the target nucleus mass, and for $N = p, n$, $Q_{\rm N}$ are the nucleon electric charges ($Q_p = 1, Q_n = 0$) and $g_{\rm N}$ are the nucleon $g$-factors ($g_p = 5.59, g_n = -3.83$). Moreover, $F_X^{\rm N, N^\prime}$ are nuclear form factors, where $X = {\rm M}, \Delta, \Sigma^\prime \Delta, \Sigma^\prime$ encompasses different nucleon properties in the target nucleus ${\rm T}$ \cite{Kavanagh:2018xeh}. The DM-nucleus differential scattering cross-section above has both spin-dependent (SD) and spin-independent (SI) interaction terms because the dark matter magnetic moment couples to the nuclear magnetic moment and coherently to the nuclear charge current in the dark matter rest frame, respectively \cite{Banks:2010eh}. Looking at the equation, the $F_{\rm M}$ gives the standard SI cross-section and the standard SD is a combination of $F_{\Sigma^\prime}$ and $F_{\Sigma^{\prime\prime}}$, in which the last one does not appear here. The reader may find richer details in \cite{Fitzpatrick:2012ix}.

\begin{figure}[t!]
    \centering
    \begin{subfigure}[b]{0.48\linewidth}
        \includegraphics[width=\linewidth]{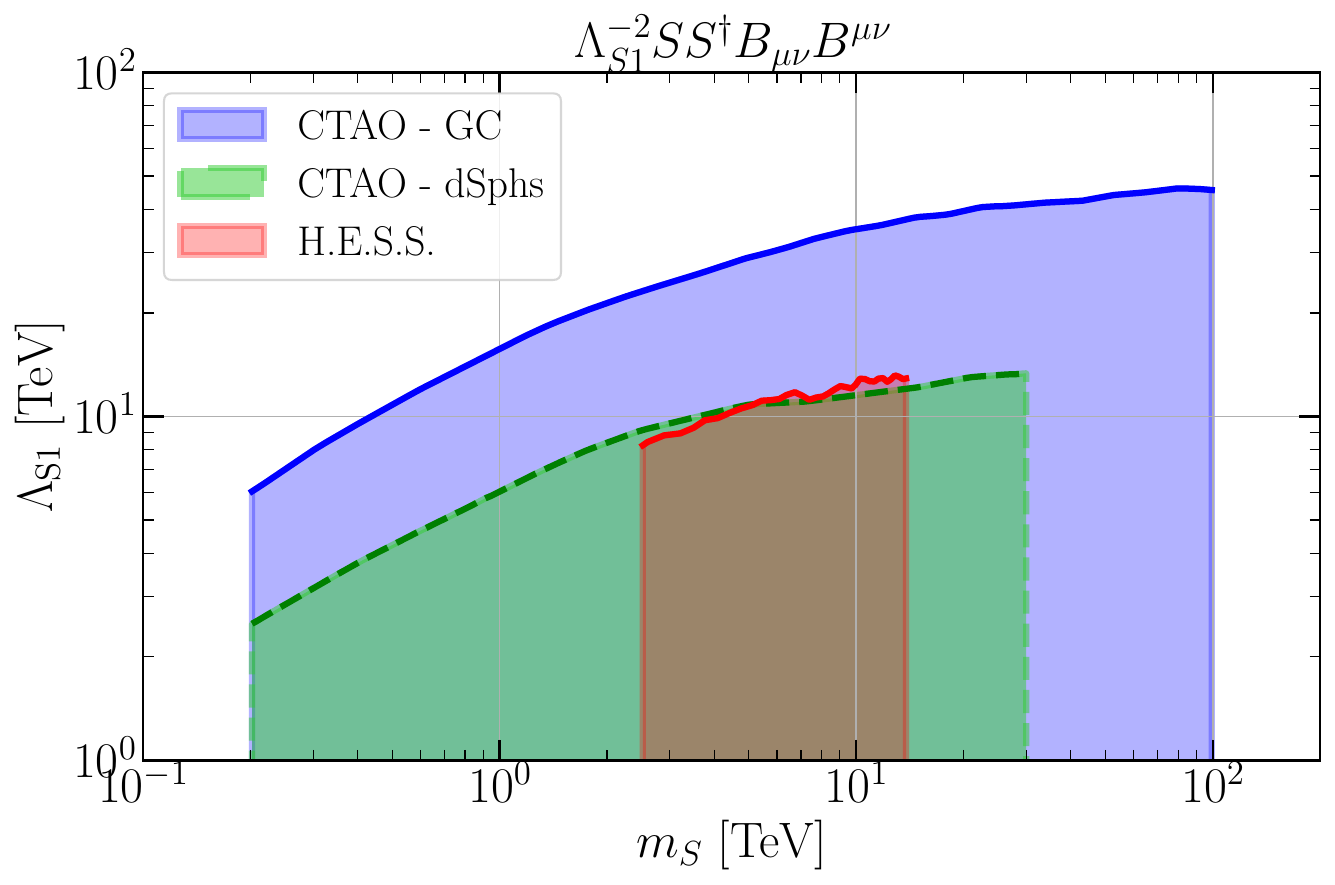}
        \caption{}
        \label{fig:S1_lines_limits}
    \end{subfigure}
    \hfill
    \begin{subfigure}[b]{0.48\linewidth}
        \includegraphics[width=\linewidth]{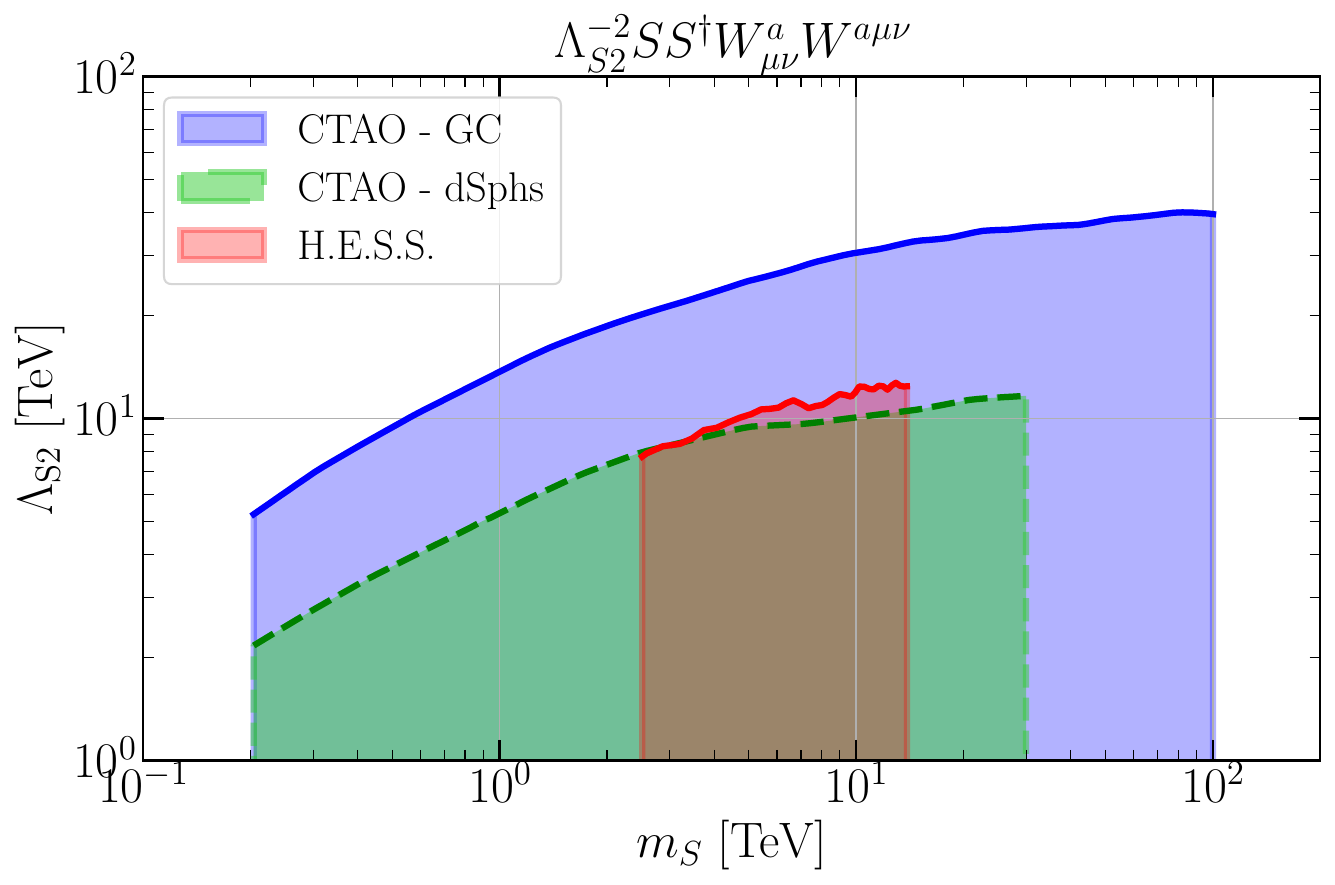}
        \caption{}
        \label{fig:S2_lines_limits}
    \end{subfigure}
    \hfill
    \begin{subfigure}[b]{0.48\linewidth}
        \includegraphics[width=\linewidth]{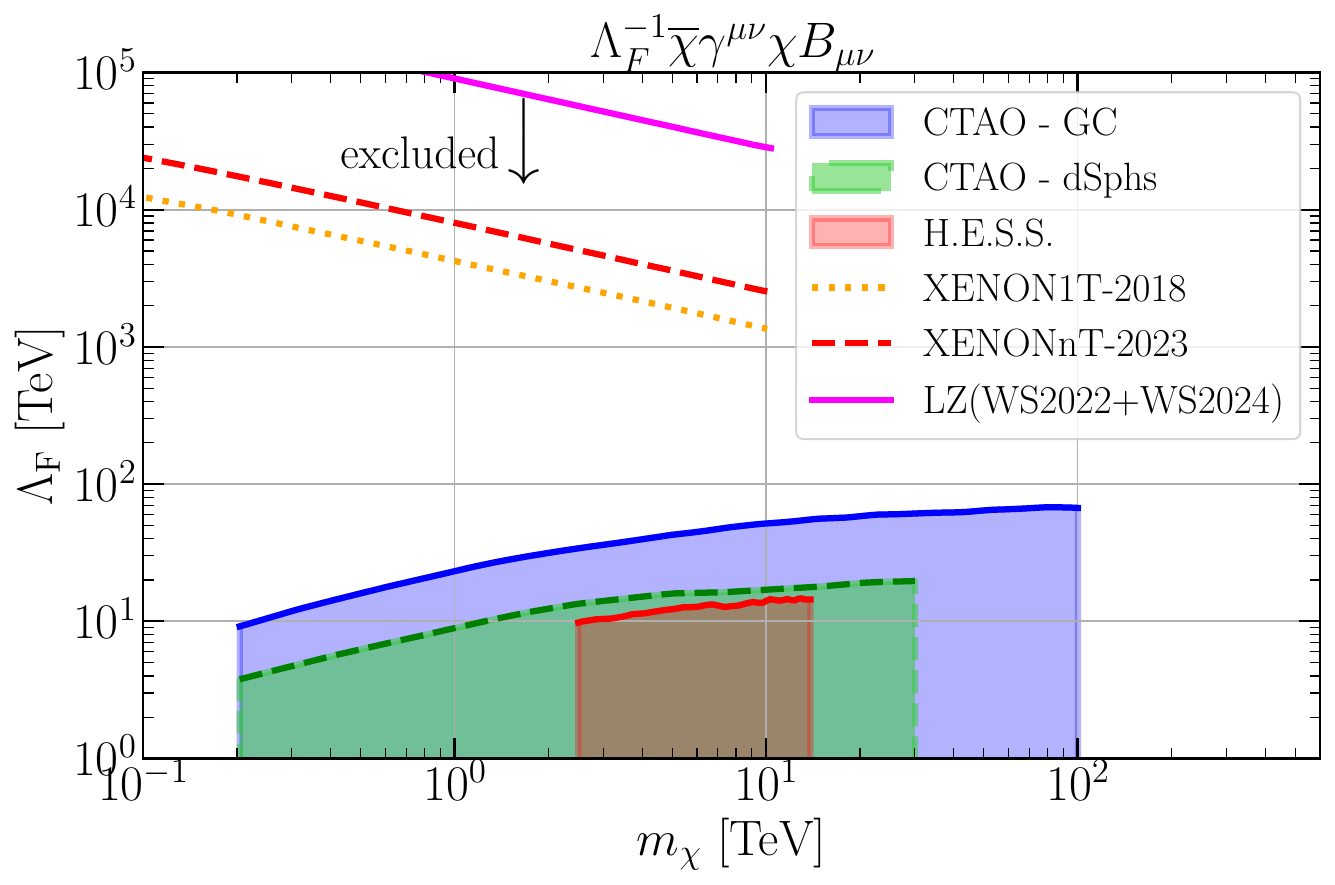}
        \caption{}
        \label{fig:F1_lines_limits}
    \end{subfigure}
    \caption{The projected lower limits on the effective energy scales as a function of the dark matter mass based on gamma-ray observations of the GC (blue) and dSphs (green).  The magenta solid (LUX-ZEPLIN), read dashed (XENONnT), and orange dotted curves (XENON1T) delimit the correct exclusion limits from direct detection experiments.}
\label{fig:line-results}
\end{figure}

Since the SD and SI interactions are tightly connected, we cannot simply compare our theoretical results against the standard SI or SD limits reported by the XENON or LUX-ZEPLIN collaborations. Therefore, we constrain the theoretical number of events predicted by the model operator, which can be determined from the differential scattering cross-section via the differential recoil rate with a given target ${\rm T}$,
\begin{eqnarray}
    \frac{dR_{\rm  T}}{dE_{\rm  R}} = \frac{\rho_\chi}{m_\chi m_{\rm T}} \int_{v > v_{\rm min}} v\, f(v) \frac{d \sigma_{\rm T}}{dE_{\rm R}} dv,
\end{eqnarray} 
where the dark matter density is $\rho_\chi$, $v_{\rm min} = \sqrt{m_{\rm T} E_{\rm R} / (2 \mu_{\chi, {\rm T}}^2)}$, with $\mu_{\chi, {\rm T}} = m_\chi m_{\rm T}/ (m_\chi + m_{\rm T})$ denoting the reduced mass, and $f(v)$ is the usual Maxwell-Boltzmann dark matter velocity distribution. Essentially, after integrating over the nuclear recoil energy taking into account the detector's efficiency, the total number of events predicted by the theory in a single energy bin is given by \cite{Cerdeno:2010jj, Cirelli:2013ufw}
\begin{eqnarray}
    N = \omega R,
\end{eqnarray}
where $\omega$ is exposure (kg$\cdot$days) and $R$ is the total event rate of DM scattering on nucleus (kg$^{-1} \cdot$ days$^{-1}$). These calculations were done in Mathematica by combining the \texttt{DirectDM} and the \texttt{DMFormFactor} software packages \cite{Bishara:2016hek,Bishara:2017nnn,Bishara:2017pfq,Brod:2017bsw,Fitzpatrick:2012ix,Anand:2013yka}. The former computed the match of Wilson coefficients of our model onto the non-relativistic EFT at nuclear scale. The latter was used to obtain the theoretical number of events because it calculates the total event rate. The total number of events was obtained from the numerical integration of the event rate and detector efficiency over the possible nuclear recoil energy. We took the efficiency curves reported by XENON1T, XENONnT, and LUX-ZEPLIN experiments \cite{XENON:2018voc,XENON:2023cxc,LZCollaboration:2024lux}. We show the approximate limits for each experiment in Fig.~\ref{fig:line-results}. The dotted orange, dashed red, and solid magenta curves are the XENON1T, XENONnT, and LUX-ZEPLIN limits that exclude the parameter space region below the curves. 

\section{Results}
\label{sec:results}
Using the Alpha configuration, the CTAO has derived expected upper limits on the dark matter annihilation cross-section into gamma-ray lines at $95\%$ confidence level (C.L.) for two astrophysical targets, the GC and dSphs  \cite{CTAO:2024wvb}. We have written effective operators for scalar and fermion dark matter particles that represent dark matter annihilation into gamma-ray lines. The dark matter annihilation cross-sections depend on the effective energy scale and dark matter mass. With this information, we can use CTAO expected limits to produce expected bounds on the effective energy scale
as a function of the dark matter masses $m_X$ for each astrophysical target. 

These results are summarized in Fig.~\ref{fig:line-results}. The solid pink and dashed green curves represent the lower limits for GC and dSphs, respectively, as the thermally averaged annihilation cross-section is inversely proportional to the energy scale. Since the $J$-factor of the GC target is much higher than that of dSphs, which increases the probability of a dark matter signal, the GC places stronger limits on the effective energy scales than the dSphs. As expected, the dSphs are less stringent but more solid because are subject to smaller astrophysical uncertainties.

Stronger experimental limits on the dark matter annihilation cross-section translate into more stringent lower bounds on the new physics effective scale. Consequently, operators predicting larger annihilation rates are more tightly constrained. This explains why operators for fermionic dark matter are more restricted than those for scalar dark matter. Similarly, the operator coupling scalar dark matter to the hypercharge field strength, $S S^{\dagger} B_{\mu\nu}B^{\mu\nu}$, is more constrained than the operator for the weak gauge fields, $S S^{\dagger} W^{a}_{\mu\nu}W^{a\mu\nu}$,  due to the difference in coupling strengths.

We observe that the CTAO is expected to probe effective scales well beyond current limits. For dark matter in the TeV mass range, projections indicate CTAO can probe $\Lambda > 10$~TeV for both scalar and fermionic candidates. The sensitivity is particularly impressive for fermionic DM, where CTAO could reach scales of $\Lambda \sim 60$~TeV for $m_X=100$~TeV. Moreover, it is clear that observations from dSphs are less stringent overall.

In Fig.~\ref{fig:line-results}, we also exhibit the current limits from H.E.S.S telescope on gamma-ray lines from dark matter annihilation, adopting an Einasto density profile \cite{Angel:2023rdd}. H.E.S.S. searched for a monoenergetic spectral line from dark matter annihilation in the $300~{\rm GeV} - 70$~TeV energy range during 10 years (254h live time)  \cite{HESS:2018cbt}. Notably, the CTAO expected sensitivity is a factor of a few better. One should bear in mind that as the annihilation cross section is proportional to $1/\Lambda^4$, thus in the $\Lambda-mass$ plane, the difference in sensitivity between the instruments is not much perceived. In terms of annihilation cross section, CTAO is expected to improve H.E.S.S. bounds by an order of magnitude for a dark matter mass of $10$~TeV.

Regarding direct direction, the first two operators yield no signal, but the fermion operator does, in the form of a magnetic dipole moment. For this reason, we display the current limits from XENON1T (dotted orange), XENONnT (dashed red), and LUX-ZEPLIN (solid magenta). They are orders of magnitude stronger than those rising from any gamma-ray instrument. We point out that this conclusion is a direct consequence of the presence of a $B_{\mu \nu}$ term. Although both constraints consider the same magnetic dipole moment operator, the annihilation cross-section features higher powers of the field strength tensor to get the $\gamma\gamma$ or $\gamma Z$ final states, and then, it scales as $1/\Lambda_{\rm F}^4$. Instead, the DM-nucleus differential scattering cross-section (and so the event rate) goes as $1/\Lambda_{\rm F}^2$. Hence, one does not need to go to higher orders to get the relevant contribution to such scattering. Anyway, we solidly conclude that CTAO will probe effective energy scales above 10~TeV and constitute a powerful dark matter probe for gamma-ray lines.

\begin{table}[t!]
\centering
\begin{tabular}{| c || c | c | c |} 
 \hline 
 \rule{0pt}{2.5ex}   
 \cellcolor{blue!20} { GC} & $\Lambda^{-2}_{\rm S1}\, SS^*B_{\mu \nu}B^{\mu \nu}$& $\Lambda^{-2}_{\rm S2} \,SS^*W^a_{\mu \nu}W^{a\mu \nu}$ & $\Lambda^{-1}_{\rm F} \, \bar{\chi} \gamma^{\mu \nu} \chi B_{\mu \nu}$\\
\hline
\rule{0pt}{3.0ex}
$m_{X}$ [TeV]& $\Lambda_{\rm min}$ [TeV]&$\Lambda_{\rm min}$ [TeV]& $\Lambda_{\rm min}$ [TeV] \\ 
\hline\hline  \rule{0pt}{2.5ex}
1 & 15.75 & 13.61 & 23.23 \\ \rule{0pt}{2.5ex}
 10 &  35.19 & 30.52  & 51.62 \\ \rule{0pt}{2.5ex}
 100 & 45.65  & 39.73 & 67.11  \\ 
\hline
 \hline
 \hline
  \rule{0pt}{2.5ex}   
 \cellcolor{blue!20} { dSphs}  & $\Lambda^{-2}_{\rm S1}\, SS^*B_{\mu \nu}B^{\mu \nu}$& $\Lambda^{-2}_{\rm S2} \,SS^*W^a_{\mu \nu}W^{a\mu \nu}$ & $\Lambda^{-1}_{\rm F} \, \bar{\chi} \gamma^{\mu \nu} \chi B_{\mu \nu}$\\
\hline
\rule{0pt}{3.0ex}
$m_{X}$ [TeV]& $\Lambda_{\rm min}$ [TeV]&$\Lambda_{\rm min}$ [TeV]& $\Lambda_{\rm min}$ [TeV] \\ 
\hline\hline  \rule{0pt}{2.5ex}
0.2 &  2.47 & 2.13 & 3.75\\ \rule{0pt}{2.5ex}
 1 & 6.03 & 5.26 & 8.94  \\ \rule{0pt}{2.5ex}
 10 &  11.59 & 10.06 & 17.01 \\ 
\hline 
\end{tabular}
\caption{\label{tab:CTAbounds} CTAO projected lower bounds on the effective energy scales $\Lambda_{\mathcal{O}}$ $(\mathcal{O}=$ S1, S2, F) as a function of the dark matter masses $m_X$ $(X=S,\chi)$ from gamma-ray lines coming from the GC and dSphs.}
\end{table}

\begin{figure*}[t!]
    \centering
    \begin{subfigure}[b]{0.48\linewidth}
         \includegraphics[width=\linewidth]{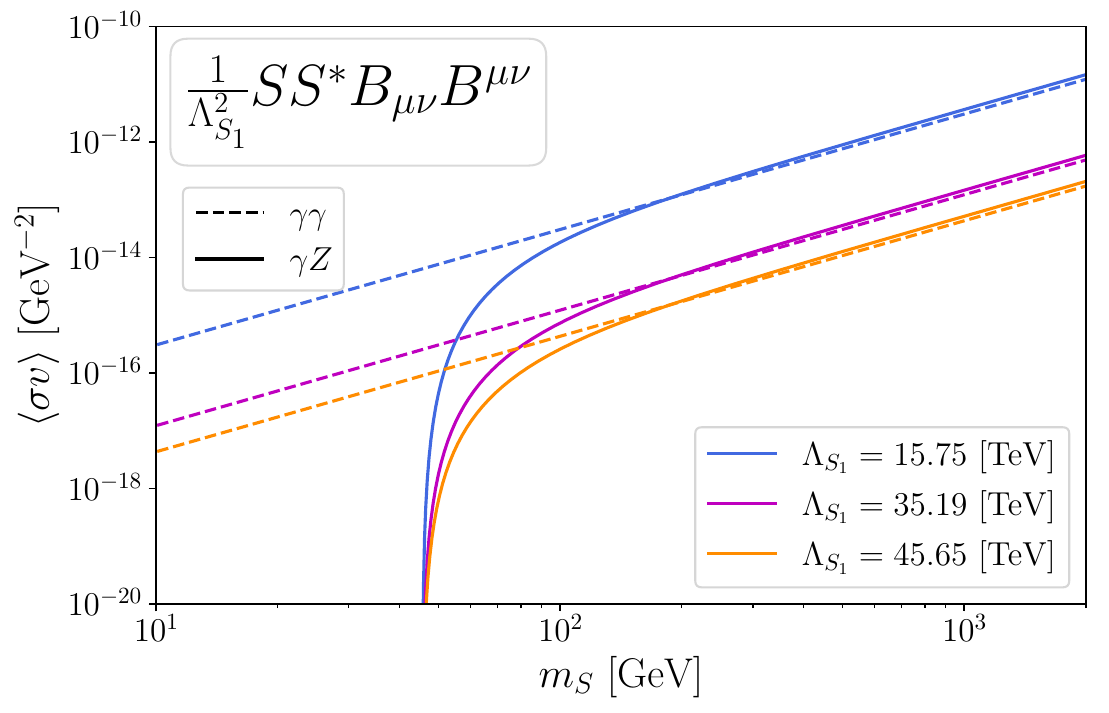}
         \caption{}
         \label{fig:sv-S1-GC}
    \end{subfigure}
    \hfill
    \begin{subfigure}[b]{0.48\linewidth}
        \includegraphics[width=\linewidth]{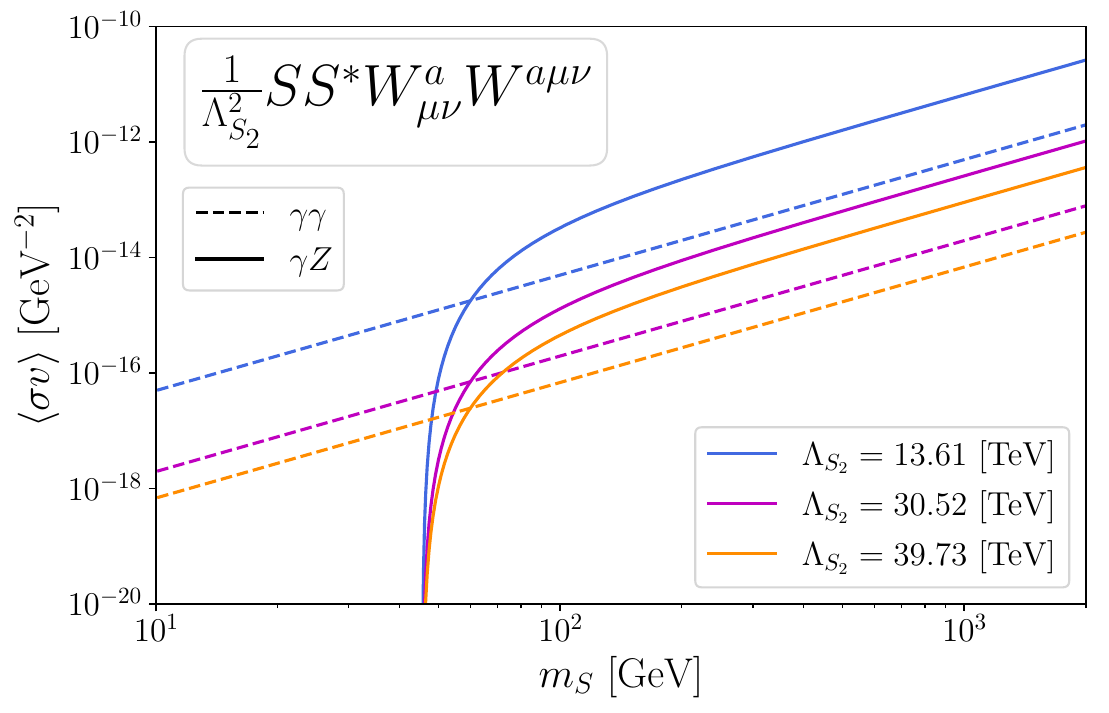}
        \caption{}
        \label{fig:sv-S2-GC}
    \end{subfigure} 
    \hfill
    \begin{subfigure}[b]{0.48\linewidth}
        \includegraphics[width=\linewidth]{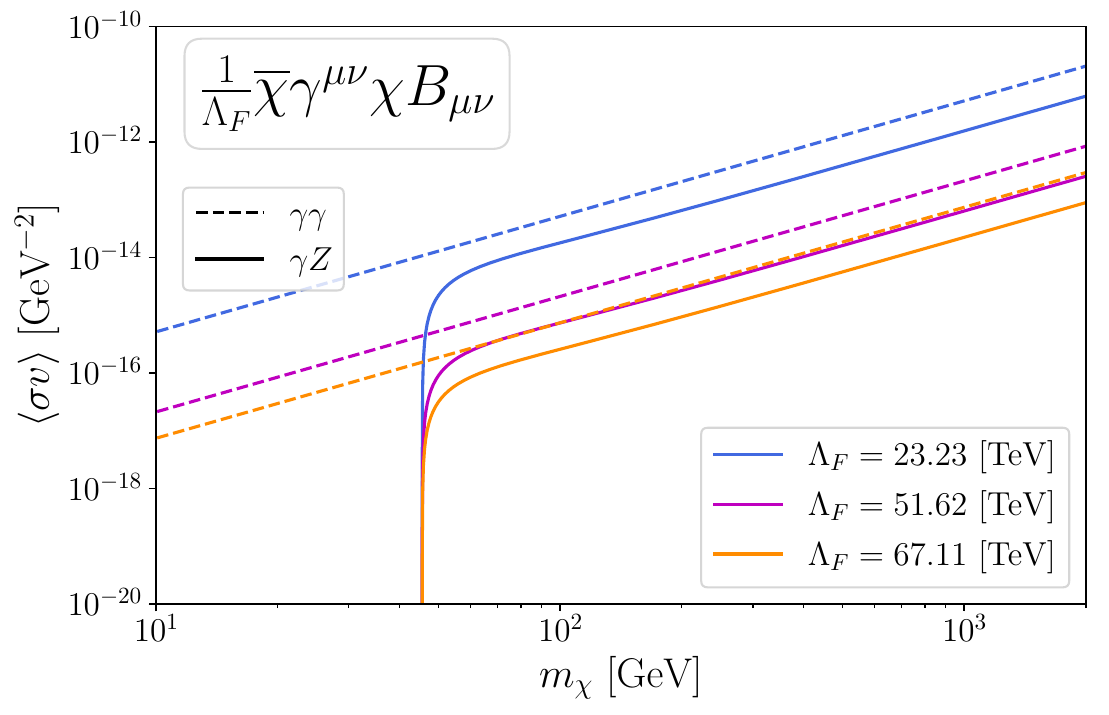}
        \caption{}
        \label{fig:sv-f-GC}
    \end{subfigure}
\caption{Galactic Centre: thermally averaged annihilation cross-section as a function of the scalar $S$ and fermion $\chi$ dark matter masses. Dark matter annihilates into $\gamma \gamma$ (dotted lines) and $\gamma Z$ (solid lines) final states for different energy scales $\Lambda$, indicated by the colors.}
\label{fig:sigma-v-GC}
\end{figure*}

\begin{figure*}
\centering
\begin{subfigure}[b]{0.48\linewidth}
  \includegraphics[width=\linewidth]{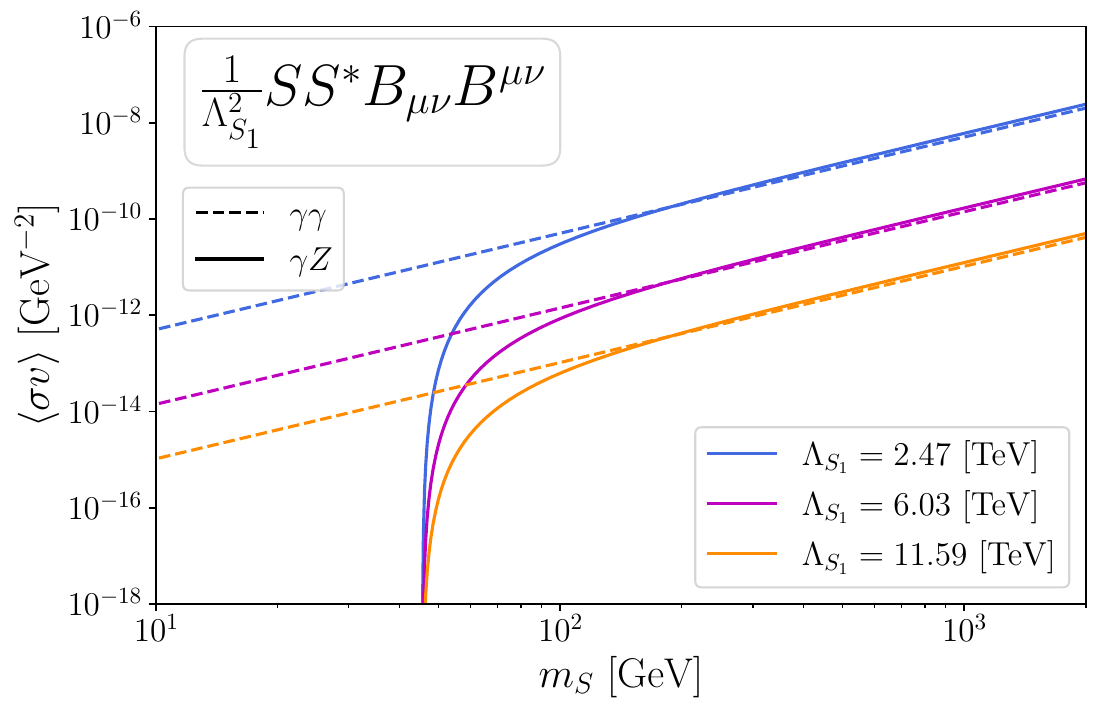}
  \caption{}
  \label{fig:sv-S1-DSph}
\end{subfigure}
\hfill
\begin{subfigure}[b]{0.48\linewidth}
  \includegraphics[width=\linewidth]{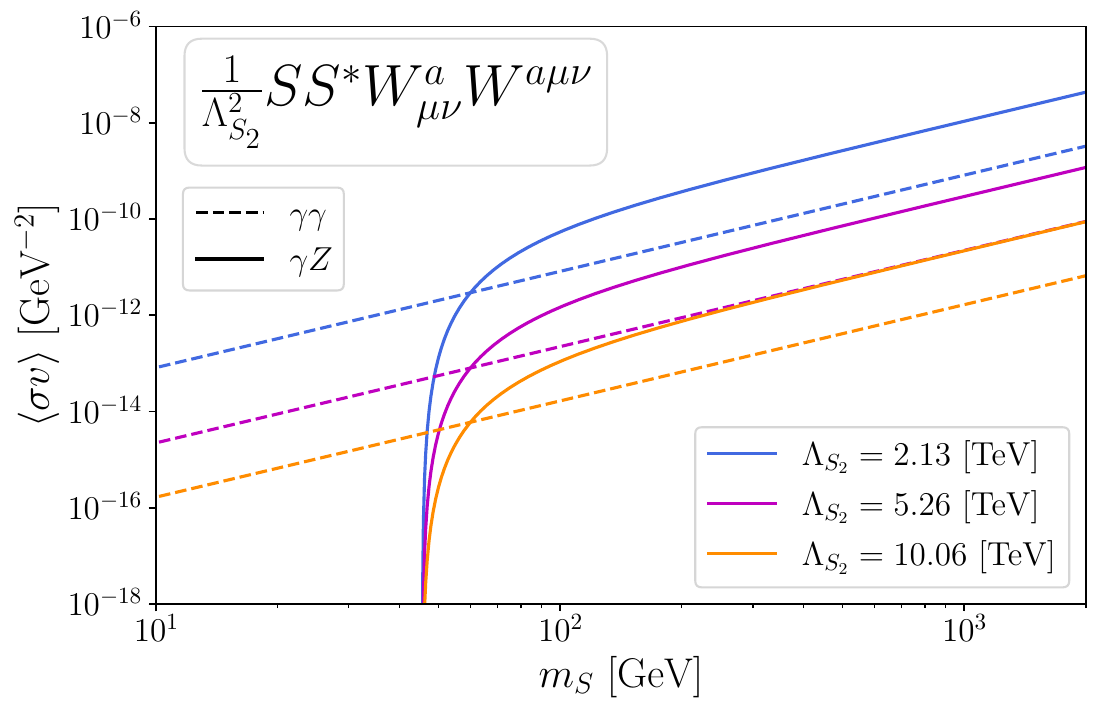}
  \caption{}
  \label{fig:sv-S2-DSph}
\end{subfigure}
\hfill
\begin{subfigure}[b]{0.48\linewidth}
  \includegraphics[width=\linewidth]{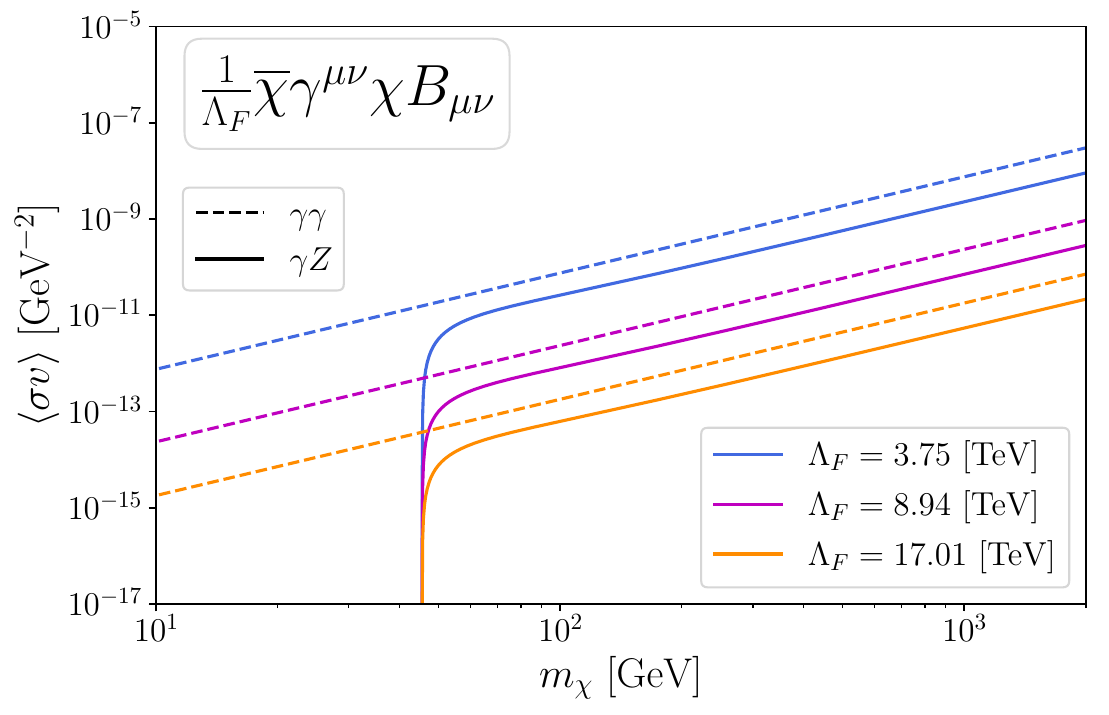}
  \caption{}
  \label{fig:sv-f-DSph}
\end{subfigure}
\caption{Dwarf Spheroidal Galaxies: thermally averaged annihilation cross-section as a function of the scalar $S$ and fermion $\chi$ dark matter masses for annihilation into $\gamma \gamma$ (dotted lines) and $\gamma Z$ (solid lines) lines for different energy scales $\Lambda$ indicated by the colors.}
\label{fig:sigma-v-targets}
\end{figure*}

We summarize the lower limits on the effective energy scales for benchmark points of $m_X$ for each astrophysical target in Tab.~\ref{tab:CTAbounds}. For completeness, we use these lower limits to plot the averaged annihilation cross-sections, from Sect.~\ref{sec:sigma}, as a function of the dark matter mass in Figs.~\ref{fig:sigma-v-GC} and \ref{fig:sigma-v-targets} for the GC and the dSphs targets, respectively. The arrangement of different sub-figures is similar: the panels (a) and (b) show the scalar dark matter annihilating into $\gamma \gamma$ and $\gamma Z$ lines through the S1 and S2 operators, respectively, with panel (c) showing the fermion dark matter annihilating into the same final states but for the fermionic operators. The colors of the curves stand for different energy scales, while the dotted and solid styles represent the $\gamma\gamma$ and $\gamma Z$ final states, respectively.

\section{Conclusions}
\label{sec:conclude}

The construction of the CTAO represents a significant advancement in our quest to comprehend the fundamental nature of dark matter. It will surpass previous gamma-ray telescopes in many ways, and probe gamma-ray photons over several decades in energy, going from $20$~GeV to $300$~TeV. Besides this broad energy coverage, the energy resolution is a key characteristic of this instrument, enabling the identification of distinct features in gamma-ray spectra associated with line signals. Having in mind that gamma-ray lines are smoking-gun signatures of dark matter annihilation and the exquisite energy resolution of CTAO at TeV energies, we assess CTAO sensitivity to scalar and fermionic dark matter using effective field theory. 

We describe dark matter annihilations into gamma-ray lines using dimension five and six effective operators and compare the expected gamma-ray signals with the projected CTAO sensitivity to gamma-ray lines stemming from the Galactic Centre and Dwarf Spheroidal Galaxies to derive projected bounds on the effective energy scale. 

Considering the emission of both $\gamma\gamma$ and $\gamma Z$ lines, our results indicate that CTAO will be able to probe effective energy scales up to $67$~TeV from gamma-ray observations of the Galactic Centre, with Dwarf Spheroidal Galaxies offering much less sensitive limits. We put our findings into perspective with direct detection data, and found that direct detection experiments can be orders of magnitude more constraining than gamma-ray instruments for the dimension five operator involving a fermion dark matter field. For the effective operators involving a scalar field, we conclude that CTAO will constitute a discovery route. 

Our conclusions rely on the lowest-order effective operators for scalar and fermion dark matter. Either way, CTAO will set new standards in the dark matter siege.

\acknowledgments

The authors thank Guillermo Gambini, Giorgio Arcadi, Juan Carlos, Tessio Melo, Sergey Kovalenko, Juri Smirnov, and José R. Alves for insightful discussions.
Furthermore, the authors thank Clarissa Siqueira, Martin White, and Ruo-Yu Shang for their relevant internal peer review of the manuscript within the CTAO collaboration.
The authors acknowledge the National Laboratory for Scientific Computing (LNCC/MCTI, Brazil) for providing HPC resources of the SDumont supercomputer (\url{http://sdumont.lncc.br}).
The authors acknowledge the use of the International Institute of Physics (IIP) cluster ``{\it bulletcluster}''.
This work was supported by Simons Foundation (Award Number:1023171-RC), 1699/24 IIF-FINEP, FAPESP Grant 2018/25225-9, 2021/01089-1, 2023/01197-4, ICTP-SAIFR FAPESP Grants 2021/14335-0, CNPq Grant 307130/2021-5, CNPq Grant 200513/2025-7, and ANID-Millennium Science Initiative Program CN2019\textunderscore044. 
LA acknowledges the support from Coordenação de Aperfeiçoamento de Pessoal de Nível Superior (CAPES) under grant 88887.827404/2023-00. The work of D.B. is supported by the Science and Engineering Research Board (SERB), Government of India grants MTR/2022/000575 and CRG/2022/000603. D.B. also acknowledges the support from the Fulbright-Nehru Academic and Professional Excellence Award 2024-25. 
JPN acknowledges support from the Programa Institucional de Internacionalização (PrInt) and the Coordenação de Aperfeiçoamento de Pessoal de Nível Superior (CAPES) under the CAPES-PrInt Grant No. 88887.912033/2023-00.
JPN is grateful to the Mainz Institute for Theoretical Physics (MITP) of the Cluster of Excellence PRISMA+ (Project ID 390831469) for its hospitality and partial support during the initial stages of this work. 
JPN thanks the University of Liverpool for the hospitality during the final stages of this project. 

\bibliographystyle{JHEPfixed}
\bibliography{references}

\end{document}